# Effects of carbon incorporation on doping state of $YBa_2Cu_3O_y$

A. Yamamoto[1,*], K. Hirose[1,2], Y. Itoh[1], T. Kakeshita[1] and S. Tajima[1]

1 Superconductivity Research Laboratory, ISTEC, 1-10-13 Shinonome, Koto-ku,

Tokyo 135-0062, JAPAN

2 Shibaura Institute of Technology, 3-9-14 Shibaura, Minato-ku, Tokyo, 116-0002 JAPAN

*: To whom correspondence should be addressed.

Present address: Musashi Institute of Technology, Advanced Research Laboratories,

8-15-1 Todoroki, Setagaya-ku, Tokyo 158-0082, Japan.

E-mail: ayamamot@sc.musashi-tech.ac.jp

Abstract

Effects of carbon incorporation on the doping state of $YBa_2Cu_3O_y$ (Y-123) were investigated. Quantitative carbon analysis revealed that carbon could be introduced into Y-123 from both the precursor and the sintering gas. Nearly carbon-free (< 200 ppm) samples were prepared from a vacuum-treated precursor by sintered at 900 ℃ and cooling with 20 ℃ /min in flowing oxygen gas. The lower $T_c$ (= 88 K) and higher oxygen content (y = 6.98) strongly suggested the overdoping state, which was supported by the temperature dependence of resisitivity and thermoelectric power. The nuclear quadrapole resonance spectra and the Raman scattering spectra indicated that there was almost no oxygen defect in the Cu-O chain in these samples. On the other hand, in the same cooling condition, the samples sintered in air stayed at optimal doping level with $T_c$ = 93 K, and the intentionally carbon-doped sample was in the underdoping state. It is revealed that about 60% of incorporated carbon was substituted for Cu at the chain site in the form of $CO_3^{2+}$, and the rest remains at the grain boundary as carbonate impurities. Such incorporation affected the oxygen absorption process in Y-123. It turned out that the oxygen content in Y-123 cannot be controlled only by the annealing temperature and the oxygen partial pressure but also by the incorporated carbon concentration.

1. INTRODUCTION

A precise control of doping level of $YBa_2Cu_3O_y$ (Y-123) is very important for advanced studies of both fundamental and application. The doping state of Y-123 can be basically varied with oxygen content. However, it is very difficult to achieve the overdoping state in polycrystalline samples by oxygen annealing, whereas the overdoped sample is easily obtained in case of single crystal Y-123 [ 1 ]. Although fully oxygenated Y-123 with y > 6.99 must be overdoped with $T_c$ = 87–89 K [ 2 ], the $T_c$ of well-oxygenated polycrystal [ 3 ] is saturated at about (=92–93 K ). The reason for this difference between the single- and the polycrystalline samples was not clarified so far.

In our preliminary experiments, we succeeded in preparing overdoped polycrystals. Key points of our method were to prepare vacuum-treated precursor and to sinter it in high-purity oxygen. This is suggestive of some effects of carbon in achieving the overdoped state.

The effects of carbon and carbon dioxide in Y-123 were studied by several groups [ 4–8 ], although no one argued the effect of carbon on the doping level. Uno *et al.* [ 4 ] reported that the calcination in vacuum is effective to reduce residual carbon concentration. The final carbon concnetration in the vacuum-treated sample is much lower (*ca.* 300 ppm) than that for the sample prepared without vacuum-calcination (*ca.* 1200 ppm). They found that $J_c$ was related to the carbon concentration, but they showed neither the $T_c$ value nor the other physical properties, including the carrier doping level. Gotor *et al.* [ 5 ] indicated that the chemical compositions of precursor and crucible are the key factors to suppress the residual carbon content. They also pointed out that, once liquid phase was formed, carbon was incorporated into Y-123 structure during crystallization.

From the interest of crystal chemistry, Karen *et al.* synthesized intentionally carbon-substituted phase $YBa_2Cu_{2.85}(CO_3)_{0.15}O_{6.73}$ [ 6 ]. The structure analysis by neutron powder diffraction revealed that $CO_3^{2+}$ group is located at the Cu chain site. As a result, a part of chain site oxygen was pushed out together with Cu, as is similar to the case of $Ba(CuO_x)_{1-y}(CO_3)_y$ (y < 0.5) [ 9 ]. It is noted that the lattice parameter, *c,* is remarkably shortened by carbonate substitution, accompanied with lowering orthorhmbicity.

In the present study, we synthesized Y-123 under various conditions, and carefully examined the relationship between residual carbon concentration and the carrier doping level. As a result, we found that in addition to the temperature and oxygen partial pressure in annealing procedure, the residual carbon concentration is an important parameter to control the doping level of Y-123. This knowledge gives a simple way to obtain almost oxygen-stoichiometric Y-123 polycrystal in the overdoped state.

## 2. EXPERIMENTAL

The samples were prepared from powders of $Y_2O_3$ (99.99%), $BaCO_3$ (99.99%), and CuO (99.99%) by the solid-state reaction. Four series of samples from A to D were synthesized by the following procedures.

Series A was calcinated and sintered in air. This is a standard way to prepare Y-123 ceramics. A mixture of the powders was calcinated in air at 860 ℃ for 48 h, and shaped to a bar, sintered in air at 930 ℃ for 24 h three times with intermediate grinding, and finally cooled in a furnace with a rate of *ca.* 20 ℃/min. Post annealing was carried out for a part of the samples under the conditions listed in Table 1.

Series B was synthesized through vacuum process. The same mixture powders as series A was calcinated in air and the bar-shaped sample was re-calcinated in a furnace with vacuum condition of *ca.* 2 x $10^{-3}$ Torr. The sample was sintered at 900 ℃ for 12 h in flowing high-purity oxygen gas in which the total carbon-including impurity (mainly CO and $CO_2$) is less than 0.05 ppm, and cooled in a furnace with the same rate of series A. A part of the samples were quenched from various temperatures listed in Table 1. Series C was synthesized to separate the effect of air calcinations and oxygen annealing, as shown in Table 2.

Series D was intentionally carbon-doped samples. They were started from vacuum-treated precursors, same as Series B, and sintered in flowing $O_2$ gas mixed with $CO_2$ (400, 800, 2000 and 4000 ppm). The $CO_2$ concentration of 400 ppm is a standard value in air.

An X-ray powder diffraction (XRD) was measured for all samples to examine the phase and purity. The lattice parameters were determined from the XRD pattern by the least squared method. Oxygen contents were determined by the iodometric titration. Metal ratio was examined by the inductively coupled plasma (ICP) method to check a probability of vaporization during sintering. Carbon concentration was analyzed by the infrared absorption method according to the regulation of Japanese Industrial Standards G1211. The values Table 1-3 were the averages of several measurements.

Dc-susceptibility of the sample was measured by a superconducting quantum interference device (SQUID) magnetometer with a magnetic field of 20 Oe in both zero-field cooling and field cooling modes. Resistivity and thermoelectric power were measured with the conventional four-probe method.

Zero-field $^{63,\,65}$Cu nuclear quadrupole resonance (NQR) spectra were measured by a spin-echo technique for the powder samples at 77.3 K in order to characterize the oxygen deficiency. The integrated spin-echo signals were recorded with changing frequency. Raman scattering measurements were performed to see the optical phonon behavior. We can roughly judge the oxygen content from the peak frequency and intensity of the apical oxygen phonon.

## 3. RESULTS AND DISCUSSION

### 3.1 Effects of suppression of residual carbon concentration

First, we examined Sample A1 that was prepared by a standard process; namely, the mixture of $Y_2O_3$, $BaCO_3$, and CuO was calcinated and sintered in air, and finally cooled naturally in a furnace (ca. 20℃/min in our furnace). All peaks of the XRD pattern are well indexed for orthorhombic cell with lattice parameters, $a$ = 3.817(1) Å, $b$ = 3.886(1) Å and $c$ = 11.678(3) Å. These are in good agreement with the reported values [ 3 ]. $T_c$ determined by the dc-susceptibility (refer to Fig. 2(a)) was 93 K that is almost the highest superconducting temperature in this material.

To introduce more oxygen, Sample A2 was post-annealed at low temperature ( 270 ℃) for a week in flowing high-purity oxygen. Another sample (A3) was cooled with a very slow rate of 0.5 ℃ /min from 900 ℃ to 200℃ in air. The $T_c$ value of both samples decreased about 1 K, suggesting a slight shift to the overdoped regime. The superconducting transition was broadened, which was probably due to oxygen inhomogeneity. High-pressure (10 atm) oxygen annealing was not effective to introduce more oxygen.

We suspected that residual carbon-included impurity might prevent introduction of oxygen. There are at least two possibilities. One is that the impurity at the grain boundary or grain surface acts as barriers of oxidation. The other is that the replacement of Cu with $CO_3^{2+}$ reduced an effective oxygen content contributing to carrier doping. In order to discuss these possibilities, we first needed to prepare a sample with a lower carbon concentration. Then, we took the Uno's method [ 4 ] and synthesized samples through a vacuum procedure.

The XRD pattern of vacuum-treated sample (B1) was shown in Fig. 1. These patters are in good agreement with Uno's report [ 4 ]. For the precursor, they assigned several peaks to unknown phase, but we found the peaks came from $BaCu_2O_2$, as indicated in the figure. The sample after sintering in pure oxygen was a single phase within a detection limit of XRD. The carbon concentration of this sample (B1) was determined to be 185 ppm that is much lower than the value (555 ppm) of Sample A1 prepared in air. ICP analysis of these samples indicated that there was no change in metal ratio before and after vacuum-treatment, which means no off-stoichiometry at 900 ℃.

Dc-susceptibility of the above two samples are shown in Fig. 2(a). $T_c$ of Sample B1 ( 88 K) is 5 K lower than that of Sample A1. Oxygen content was 6.86 and 6.98 for Sample A1 and Sample B1, respectively. This indicates that the vacuum-treated sample (B1) is very close to stoichiometric composition of $YBa_2Cu_3O_7$. To ensure the doping level of Sample B1, the oxygen content was reduced by quenching from 400–900 ℃, as listed in Table 1. Figure 2(b) showed the dc-susceptibility of the quenched samples (B2 and B3). As rising the quenching temperature, $T_c$ increased and had a maximum at 93.5 K in the sample quenched from 460 ℃ ,

and further rising temperature led to a decrease in $T_c$. These results proved that Sample B1 was in the overdoped state without any post-annealing.

So far it has been believed that long time and low temperature annealing in oxygen were needed to obtain an overdoped sample. Here we found a simple way to synthesize overdoped polycrystal with high reproducibility. At this step we cannot specify where the carbon impurity come from, and what is the effects of the impurity. This will be discussed in section 3.3

3.2 Confirmation of overdoping

Physical properties of Sample B1 were measured to find further evidence of overdoping.

Temperature dependences of electrical resistivity for Sample B1 and two optimally doped samples (B3 and A1), are plotted in Fig. 3(a). $T_c$ of B1 is lower than those of B3 and A1. The $\rho$-$T$ curve of B1 keeps a good linearity down to $T_c$, while those of B3 and A1 slightly drop from 30-50 K above $T_c$, which is consistent with the results of single crystals [ 10 ]. Though the absolute values of resistivity are considerably higher than those of single crystal, we supposed that it is an effect of grain boundaries.

Figure 3(b) showed temperature dependence of Seebeck coefficient ($S$). The $S$-$T$ curve with the room temperature value of $S = -3\mu V/K$ for B1 shows a good agreement with that of overdoped single crystal ($T_c$ = 87 K, y = 6.99) as well as the absolute value at room temperature [ 11 ]. The $S$-$T$ curves for the optimally doped samples (B3 and A1) are also the same as those of nearly optimally doped single crystal [ 11 ].

To examine a degree of oxygen deficiency in the CuO chain, the zero-field $^{63,65}$Cu NQR frequency spectra were measured for the sample B1, B3 and A1 at 77.3 K. Figure 4 show the respective Cu(1) (CuO chain) and Cu(2) ($CuO_2$ plane) NQR spectra for B1 and B3. Very sharp peaks in the spectrum for B1 indicate no appreciable oxygen deficiency, while broad and tailed peaks in B3 suggest charge distribution due to oxygen deficienciy [12]. The spectrum of A1 was almost the same as that of B3. The $^{63}$Cu nuclear spin-lattice relaxation curves also confirmed that the sample B1 has a stoichiometric composition in oxygen, but A1 and B3 have some defects.

Raman spectra for series B are shown in Fig. 5. The spectrum for the sample quenched from 600 ℃ (B5) is a typical one for oxygen deficient Y-123 [ 13 ] . The peak near 500 $cm^{-1}$ is ascribed to stretching mode phonon for the apical oxygen, while the peak at about 580 $cm^{-1}$ corresponds to the same mode but for the chain oxygen deficient sites. As the oxygen content increases, the 500 $cm^{-1}$ peak shifts to higher wave number which indicates shortening of the bond distance between Cu(2) and the apical oxygen. Simultaneously, the defect-induced mode at about 580 $cm^{-1}$ vanishes.

These results demonstrates that the oxygen deficiency in Sample B1 is clearly lower than that of optimally doped samples (B3 and A1), and the oxygen content of B1 is comparable to that of low-temperature annealed single crystals.

3.3 Carbon source and location

The first question is what are the carbon sources. To clarify it, we prepared additional samples under various synthetic conditions, as listed in Table 2.

Sample C1 was prepared by skipping the vacuum treatment, and it was sintered under the same condition as B1 to test a roll of vacuum calcination. The $T_c$ of C1 was 3.5 K higher than that of B1(See Table 2 and Fig. 6), and the residual carbon concentration of C1 is much higher than that of B1. This demonstrates that one of the carbon impurity sources is residual carbon in the precursor. It was supported by the direct carbon analysis for the precursor. The carbon concentration of the air-heated precursor (2031 ppm) was five times larger than that of vacuum-heated precursor (413 ppm).

Sample C2 was synthesized to check the influence of sintering gas. We used vacuum-treated precursor but sintered it in a $CO_2$-$O_2$ mixed gas in which the $CO_2$ concentration is the same as in a standard air (400 ppm). The analyzed carbon concentration of Sample C2 (638 ppm) was much higher than that of B1 (See Table 2). Correspondingly, the $T_c$ of C2 was higher than that of B1, but comparable to C1, suggesting the lower oxygen content than B1. The high carbon concentration indicates that the carbon impurity can be easily taken into the sample from gas during the sintering process. Sample C3 that was prepared without vacuum-treatment and sintered in $CO_2$ mixed gas showed almost the same $T_c$ as C1and C2, but its superconductivity volume fraction decreased dramatically, as shown in Fig. 6.

These results show that the main sources of carbon impurity in Y-123 are the precursor and the sintering gas. Absorption of $CO_2$ during sintering was also confirmed by thermogravimetric analysis of Y-123 in flowing $CO_2$-$O_2$ gas.

The next question is the location of carbon. We prepared intentionally carbon-doped samples to know the effects of carbon substitution on the Y-123 structure. As is seen in Table 3, the carbon concentration in the sample became higher with increasing carbon concentration in the gas, which suggests carbon absorption from the sintering gas.

Lattice parameter *c* was systematically shortened with carbon doping, as shown in Table 3. Here, we can estimate the substituted carbon content, based on the Karen's data obtained by neutron diffraction analysis [ 6 ]. According to their analysis, 15% replacement of chain site Cu with $CO_3^{2+}$ induces a reduction of 0.069 (1) Å in the lattice parameter *c*. They also indicated that the total carbon concentration in the sample is higher than the substituted carbon, which means there are additional impurities at the grain boundaries and/or the grain surface of Y-123. Assuming a linear relationship between the substituted carbon concentration and the reduction in *c*, we estimated the former from the latter data in our samples, as illustrated in Fig. 7 and in Table 3. This estimation suggests that about 1% of Cu(1) was replaced with $CO_3^{2+}$ in the case of conventionally synthesized Y-123.

Heavy carbon substitution makes a dramatic effect on the physical properties. Figure 8 shows the temperature dependence of susceptibility for Sample C1, D1, D2, and D3. With increasing carbon concentration, Tc was remarkably lowered and the volume fraction was reduced. The electrical resistivity and the thermoelectric power of the heavily carbon-substituted sample (D2) are shown in Fig. 9 together with those of optimally (A1) and overdoped (B1) samples. The Seebeck coefficient of Sample D2 is comparable to that of the 600℃-quenched sample (B5). Both data indicated that the carbon substitution reduces the carrier doping level, which is probably caused by insufficient CuO chain-site oxygen that acts as carrier supplier. A similar observation was reported in $Y_{0.8}Ca_{0.2}Ba_2Cu_{1-x}C_xO_y$ [11]. Therefore, it is concluded that the main reason for the Tc-reduction by carbon substitution is the decrease in carrier doping level, although we cannot ignore the effect of local structural change such as lowering orthorhombicity and a positional shift of apical oxygen accompanied with a shift of Ba [ 6 ].

## 4. CONCLUDING REMARKS

Our motivation of this work was to understand why Y-123 polycrystals do not show obvious overdoping characters, although the overdoped state was well studied for single crystals, in other words, why optimally doped sample can be easily obtained just by calcination and sintering in air. The answer of these questions is that a polycrystalline sample of Y-123 prepared in a conventional way contains a small amount of carbon, which makes its carrier doping level optimum in spite of the oxygen annealing that should give an overdoped state.

We also demonstrated that carbon came from both the precursor and the sintering gas. Reducing carbon concentration in the precursor and the sintering gas leads to nearly carbon-free and oxygen-stoichiometric Y-123. Furthermore, the amount of substituted carbon in Cu was estimated from the shrinking of lattice parameter *c*. According to our analyses, the conventionally prepared Y-123 has about 1% carbon-substitution at the Cu chain site and oxygen cannot be fully introduced, resulting in the optimally doping state. These results require us to reexamine the complex and partly contradictory results reported for slightly overdoed polycrystals.

To understand the whole effects of the carbon incorporation in Y-123, it is necessary to inspect carbonate impurities at the grain boundary, compositional restructuring at the grain surface [ 7 ], or localization of substituted carbon along the twin boundary [ 8 ], and so on. These impurities seem to affect a diffusion rate of oxygen, and it is one of the most important practical problems to control the doping state in application products. Since the effects may be correlated with each other, it is not easy to figure out completely.

Finally, we would like to mention that the carbon incorporation into the samples is not a problem only for Y-123 but also for most of high-temperature superconducting (HTSC) copper oxides, especially for the Ba-including HTSCs obtained by sintering in air or in the sealed tube.

ACKNOWLEDGMENTS

We greatly appreciate Dr. N. Chikumoto for measurements of the electrical resisitivity. We also thank Drs. Masui and Uchiyama for the useful discussion. This work was supported by New Energy and Industrial Technology Development Organization (NEDO) as Collaborative Research and Development of Fundamental Technologies for Superconductivity Applications.

Table 1 Synthetic conditions, $T_c$, oxygen content and carbon concentration of $YBa_2Cu_3O_y$ prepared under various conditions

| Sample label | Synthetic conditions<br>calcination / sintering / post-annealing | $T_c$ (K) | oxygen content, y | Carbon concent-ration (ppm) |
|---|---|---|---|---|
| A1 | air, 860˚C, 48h / air, 930˚C, 12h x 3 [1]FC / no annealing | 93 | 6.86 | 555 |
| A2 | air, 860˚C, 48h / air, 930˚C, 12h x 3 FC / $O_2$, 270˚C, 1week, FC | 92 | 6.88 | |
| A3 | air, 860˚C, 48h / air, 930˚C, 12h x 3 FC / air, 930˚C, 12h, 0.5˚C/min | 91 | | |
| B1 | vac., 900˚C, 12h / $O_2$, 900˚C, 12h, FC / no annealing | 88 | 6.98 | 185 |
| B2 | vac., 900˚C, 12h / $O_2$, 900˚C, 12h, FC / $O_2$, 400˚C, quench into Liq. $N_2$ | 91 | 6.94 | |
| B3 | vac., 900˚C, 12h / $O_2$, 900˚C, 12h, FC / $O_2$, 460˚C, quench into Liq. $N_2$ | 93.5 | 6.88 | |
| B4 | vac., 900˚C, 12h / $O_2$, 900˚C, 12h, FC / $O_2$, 510˚C, quench into Liq. $N_2$ | 92.5 | | |
| B5 | vac., 900˚C, 12h / $O_2$, 900˚C, 12h, FC / $O_2$, 600˚C, quench into Liq. $N_2$ | 69 | | |
| B6 | vac., 900˚C, 12h / $O_2$, 900˚C, 12h, FC / $O_2$, 700˚C, quench into Liq. $N_2$ | 56 | | |
| B7 | vac., 900˚C, 12h / $O_2$, 900˚C, 12h, FC / $O_2$, 800˚C, quench into Liq. $N_2$ | > 20 | | |

1)FC: Furnace cooing, *ca*, 20 ˚C/min in our furnace

Table 2 Synthetic conditions, $T_c$, and carbon content of $YBa_2Cu_3O_y$

| Sample label | calcination / sintering and cooling processes | $T_c$ (K) | Carbon concentration (ppm) |
|---|---|---|---|
| B1 | vac., 900˚C, 12h / $O_2$, 900˚C, 12h, FC | 88 | 185 |
| C1 | Air. 860˚C, 48h/ $O_2$, 900˚C, 12h, FC | 91.5 | 549 |
| C2 | vac., 900˚C, 12h / $CO_2$(200ppm)-$O_2$(base), 900˚C, 12h, FC | 92 | 638 |
| C3 | Air. 860˚C, 48h / $CO_2$(200ppm)-$O_2$(base), 900˚C, 12h, FC | 92 | 1663 |

Table 3 $T_c$, lattice parameters, total and substituted carbon concentrations in $YBa_2Cu_3O_y$ prepared from vacuum-treated precursor

| Sample label | $CO_2$ concentration in sintering gas (ppm) | $T_c$ (K) | Lattice parameters *a* (Å) | *b* (Å) | *c* (Å) | Total carbon concentration (ppm) | Substituted carbon concentration (ppm) | Carbon content, z, in $YBa_2(CuO_x)_{3-z}(CO_3)_z$ |
|---|---|---|---|---|---|---|---|---|
| B1 | < 0.5 | 88 | 3.817(1) | 3.884(1) | 11.681(2) | 185 | 0 | 0 |
| C2 | 400 | 93 | 3.820(1) | 3.885(1) | 11.676(2) | 638 | 190 | 0.01 |
| D1 | 800 | 92 | 3.823(1) | 3.888(1) | 11.670(2) | 1015 | 400 | 0.02 |
| D2 | 2000 | 80 | 3.827(1) | 3.887(1) | 11.663(3) | 1731 | 670 | 0.04 |
| D3 | 4000 | 65 | 3.828(1) | 3.889(1) | 11.640(3) | 2581 | 1500 | 0.08 |
| R1[1)] | – | – | 3.8717(3) | 3.8717(3) | 11.607(1) | 4100 | 2720 | 0.15 |

1) Karen *et al*. Ref. 6

Figure captions

Fig. 1 X-ray powder diffraction patterns for $YBa_2Cu_3O_y$ of Series B (a) after calcination in vacuum at 900℃ for 12h, and (b )after sintering in high-purity oxygen at 900℃ for 12hwith furnace cooling .

Fig. 2 DC-susceptibility of $YBa_2Cu_3O_y$ samples. (a) conventional (A1) and vacuum-treated samples(B1), (b) vacuum-treated samples; furnace cooled (B1) and quenched from 400 ℃ (B2) and 460℃ (B3).

Fig. 3 Transport properties of $YBa_2Cu_3O_y$ samples. (a) Electrical resistivity and (b) thermoelectric power.

Fig. 4  Zero-field Cu NQR frequency spectra of $YBa_2Cu_3O_y$ samples at 77.3 K. The solid and dash curves are simulation with Gaussian functions.

Fig. 5 Raman scattering spectra of $YBa_2Cu_3O_y$ samples.

Fig. 6 DC-susceptibility of $YBa_2Cu_3O_y$ samples prepared under various conditions.

Fig. 7 Relationship between carbon concentration and the lattice parameter *c*. Solid and open symbols indicate the analyzed values and open the estimated values from solid line. The line is drawn, assuming no substituted carbon in Sample B1 as well as a liner relationship of substituted carbon concentration and the lattice parameter *c*.

Fig. 8 DC-susceptibility of $YBa_2Cu_3O_y$ samples sintered in various concentrations of $CO_2$ –$O_2$ gas.

Fig. 9 Transport properties of $YBa_2Cu_3O_y$ samples sintered in various concentrations of $CO_2$ –$O_2$ gas. (a)Electrical resistivity and (b) thermoelectric power.

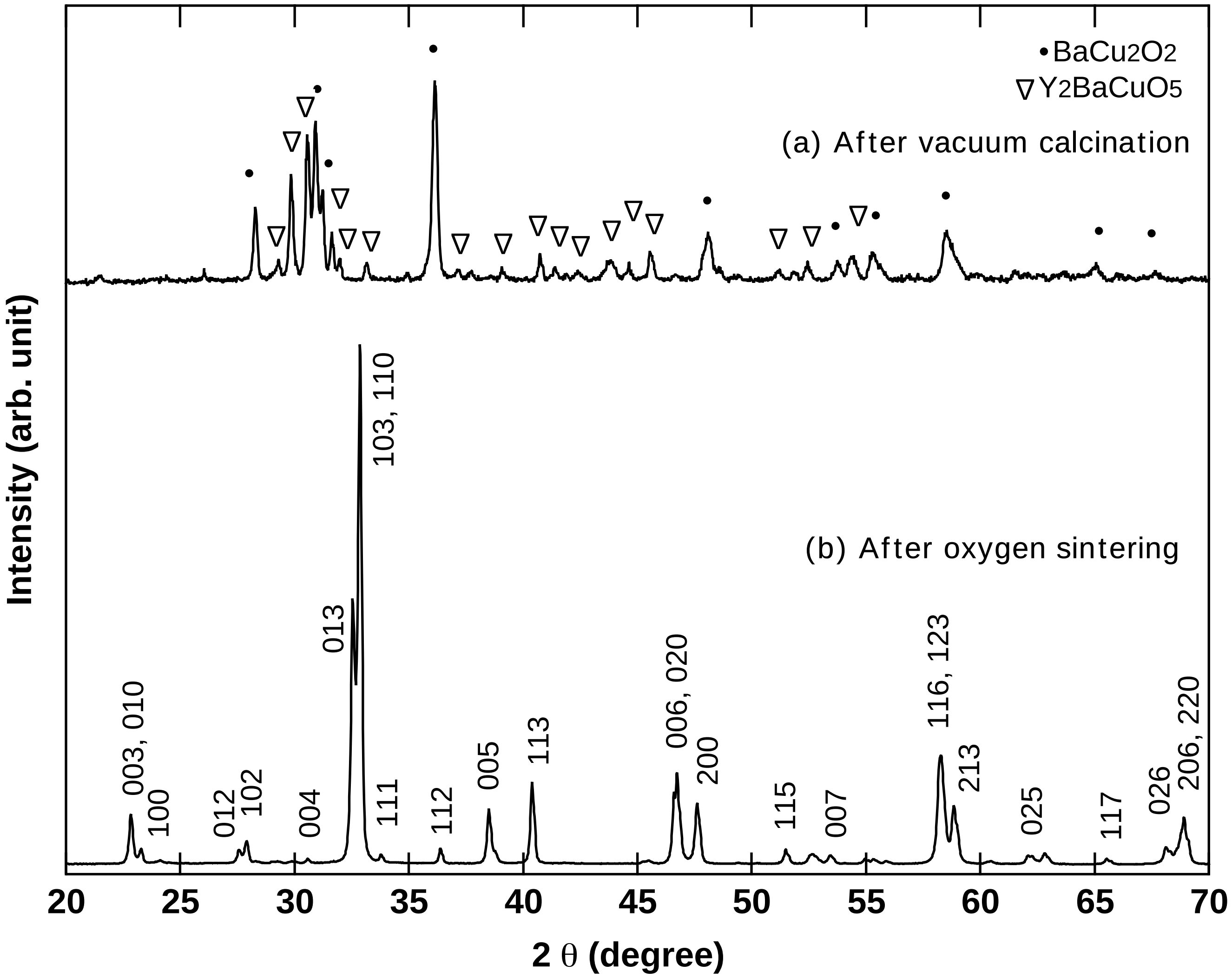

Fig. 1 A. Yamamoto et al.

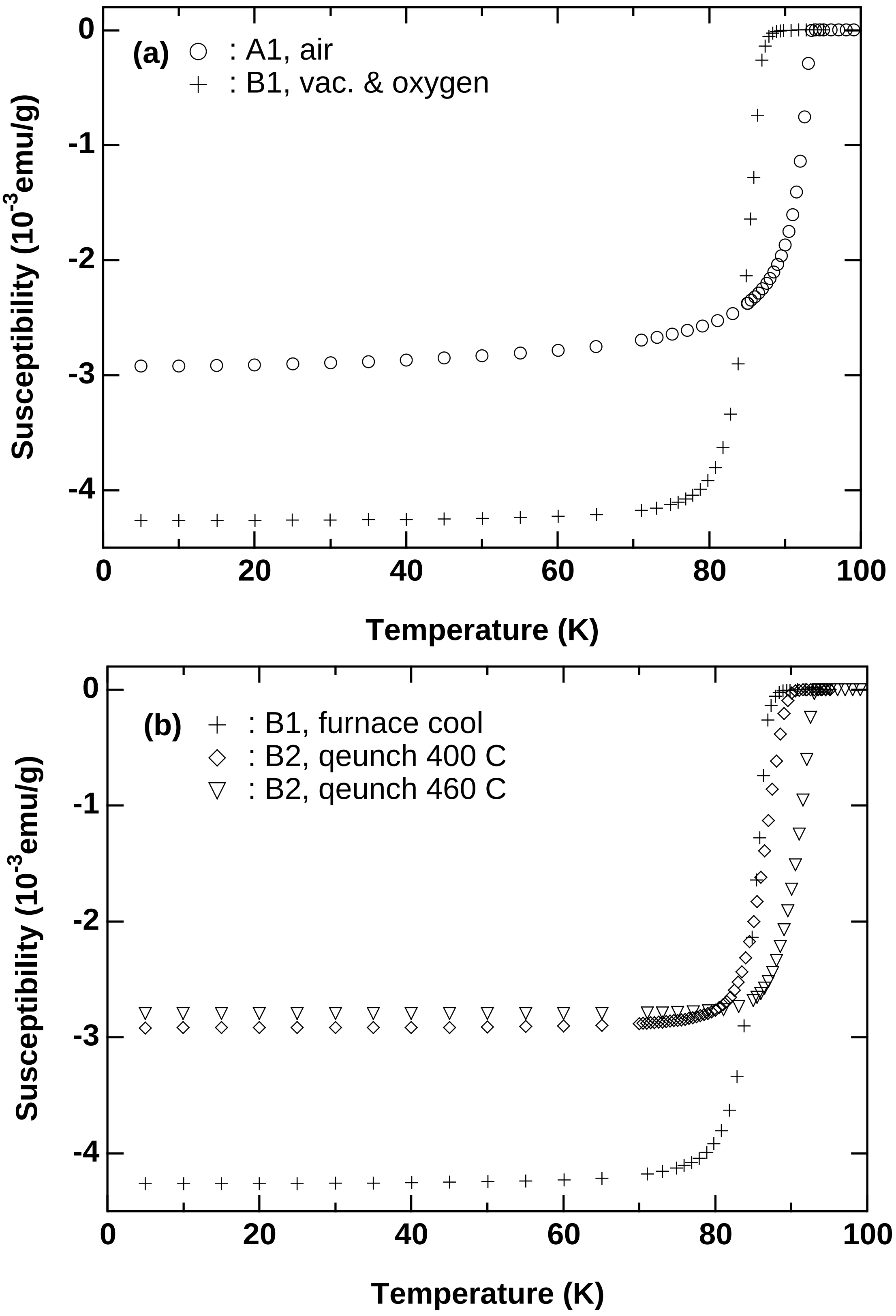

Fig. 2 A.Yamamoto et al.

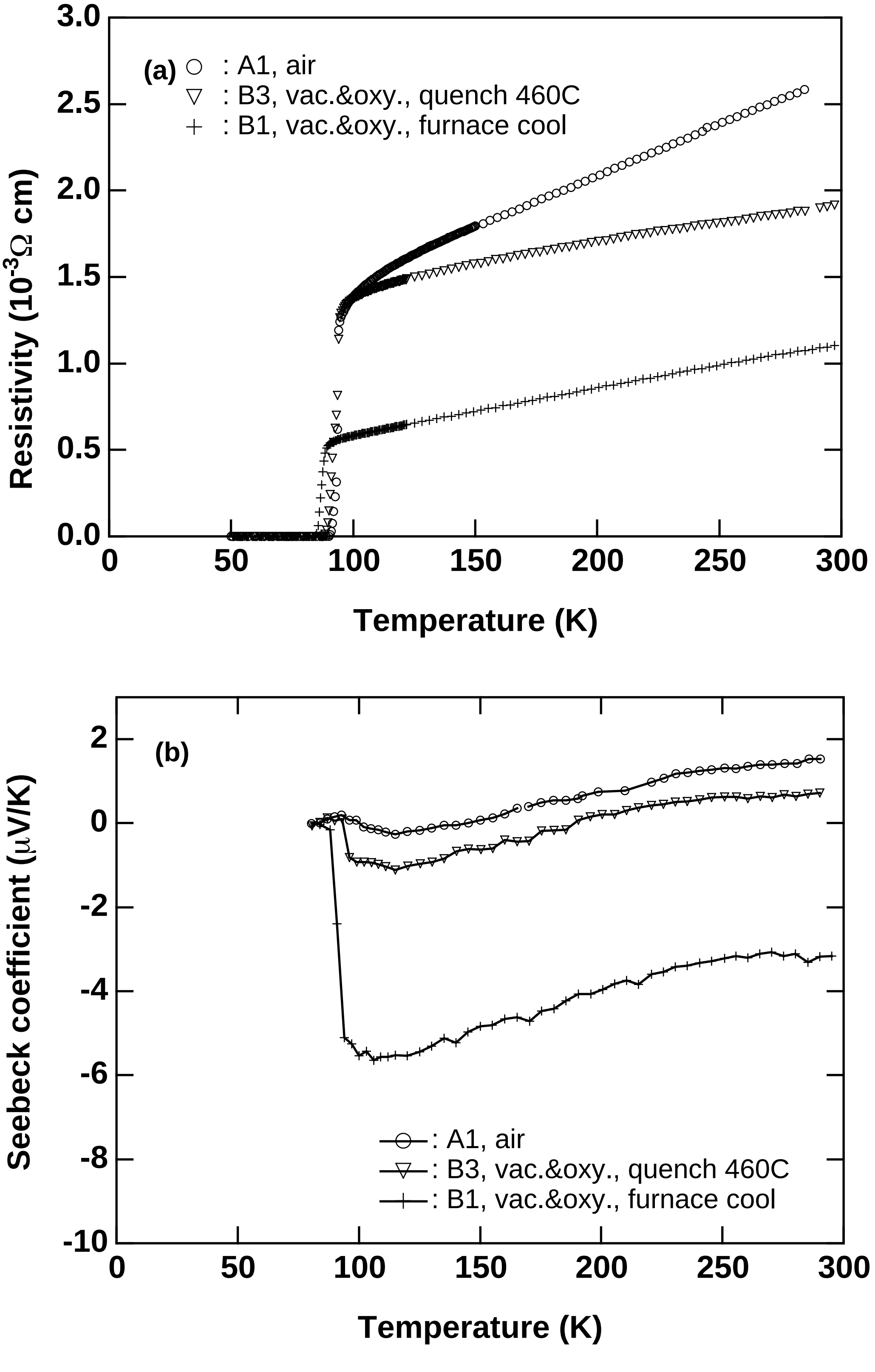

Fig. 3 A.Yamamoto et al.

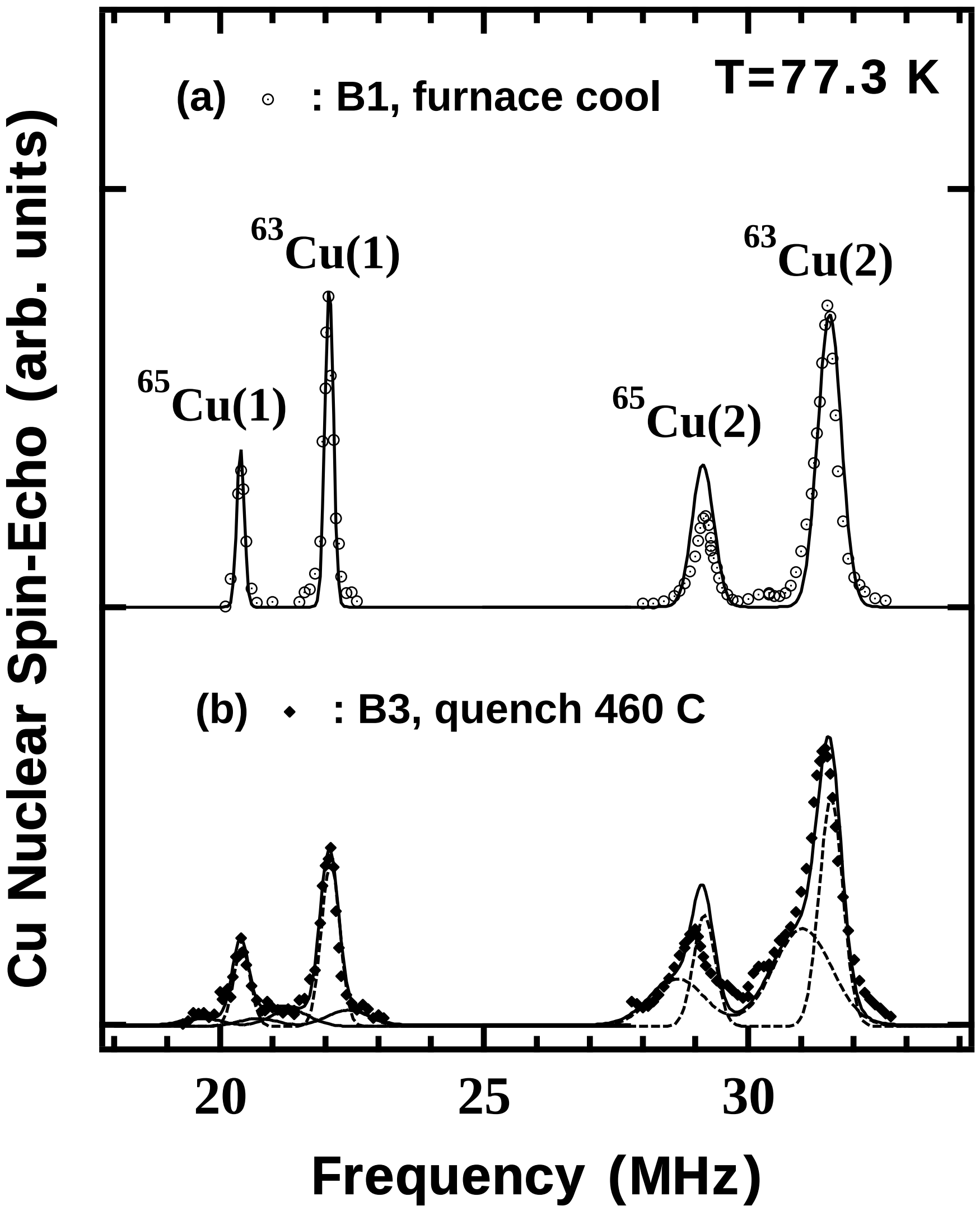

Fig. 4 A. Yamamoto et al.

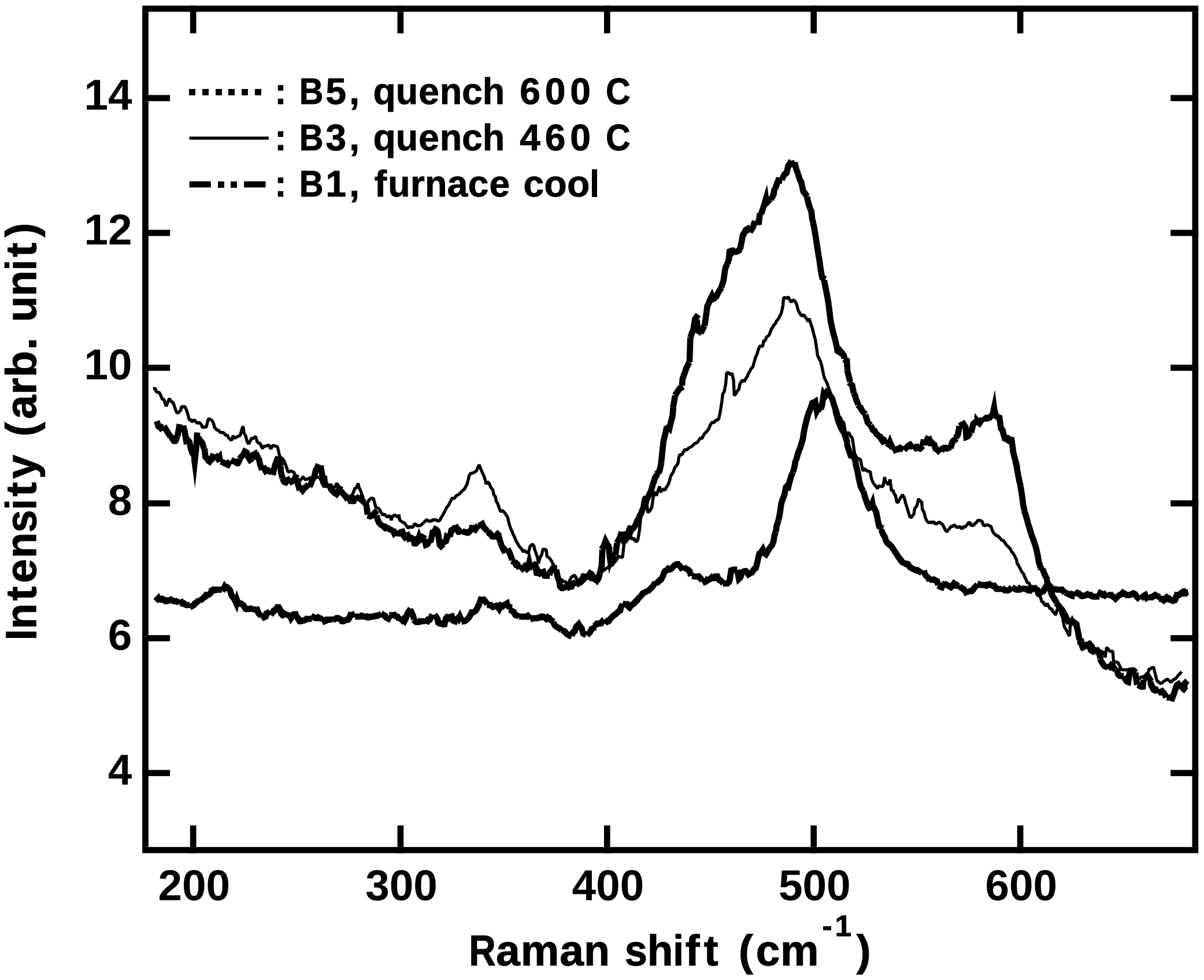

Fig.5 A.Yamamoto et al.

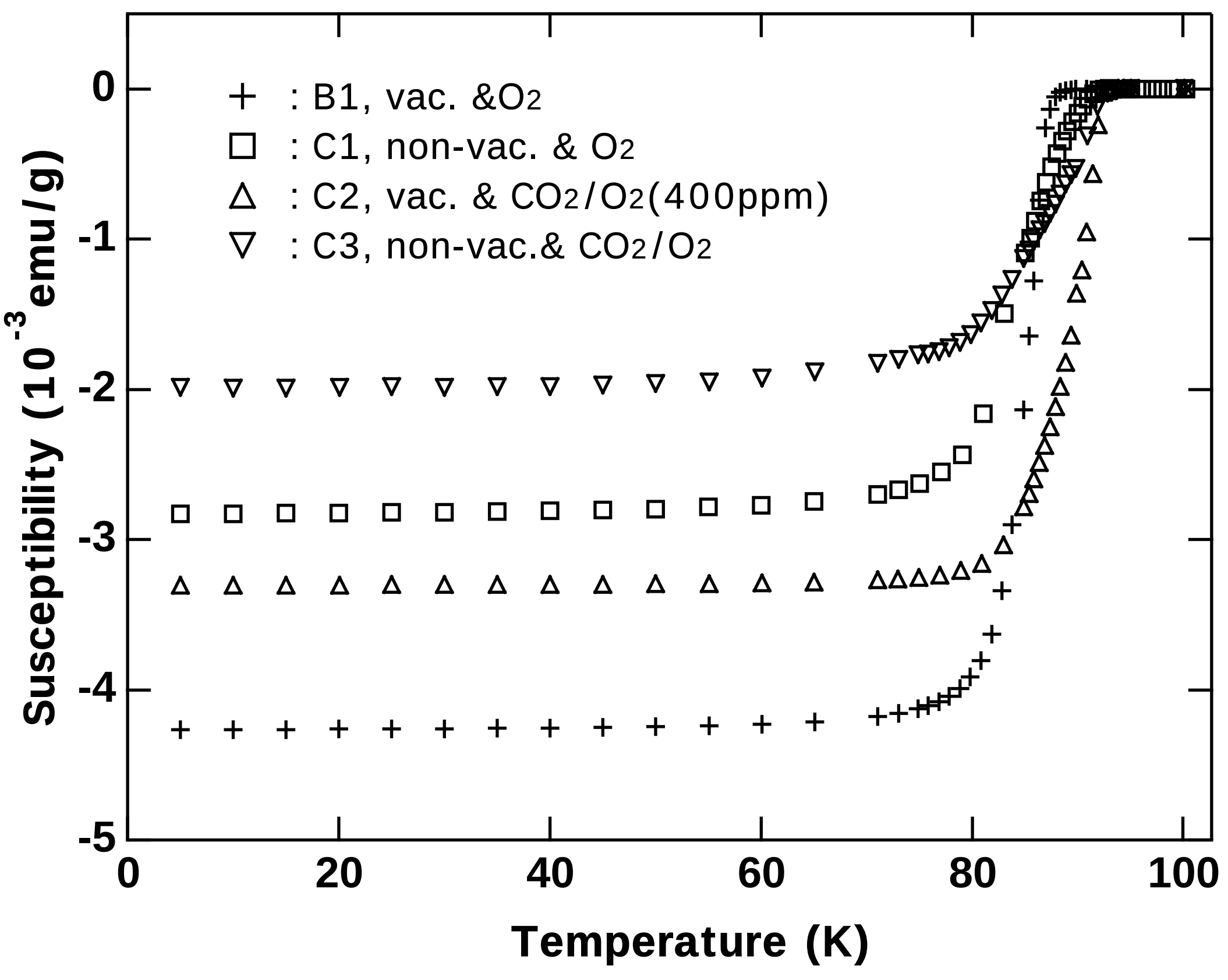

Fig. 6     A.Yamamoto et al.

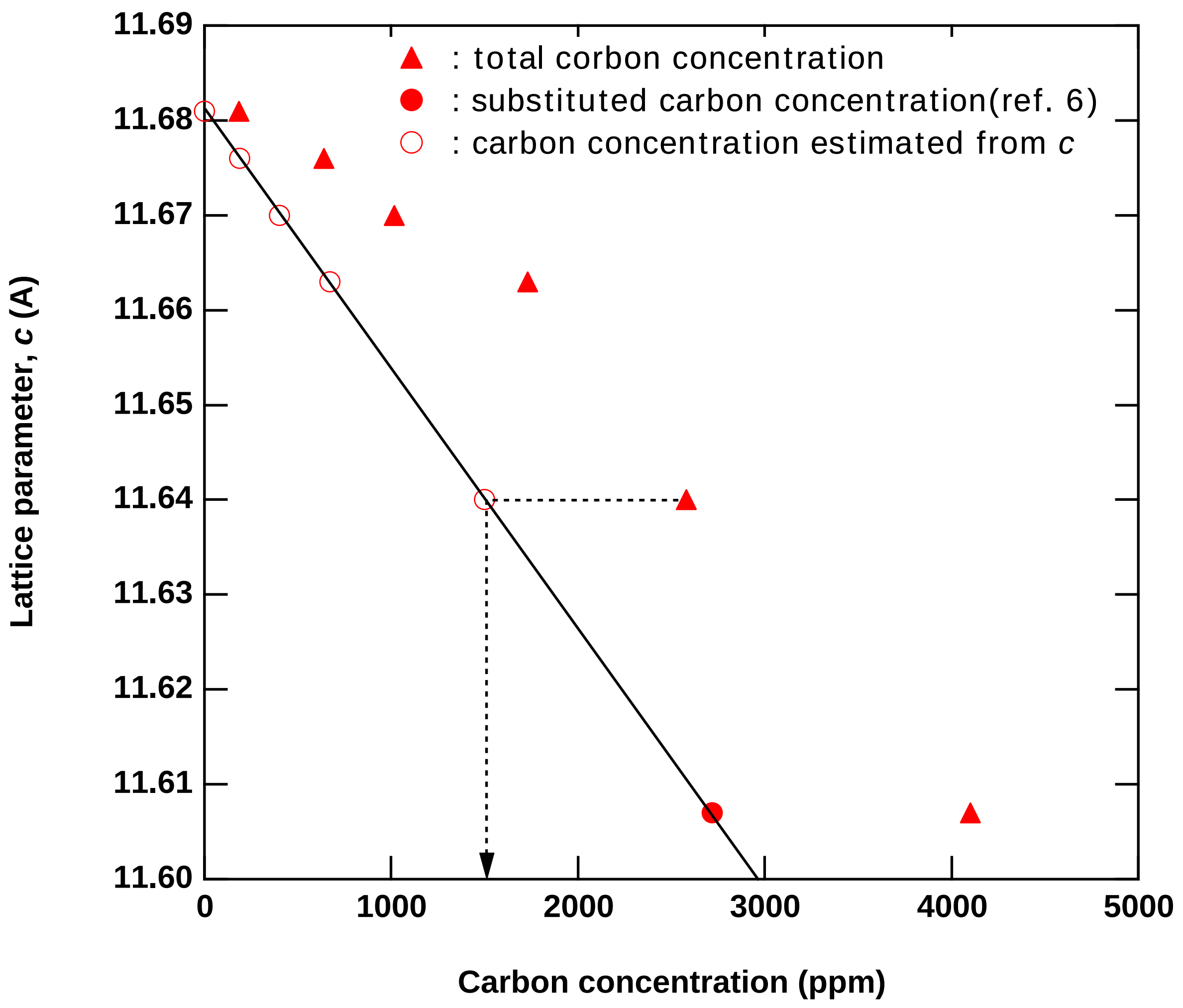

Fig. 7 A. Yamamoto et al.

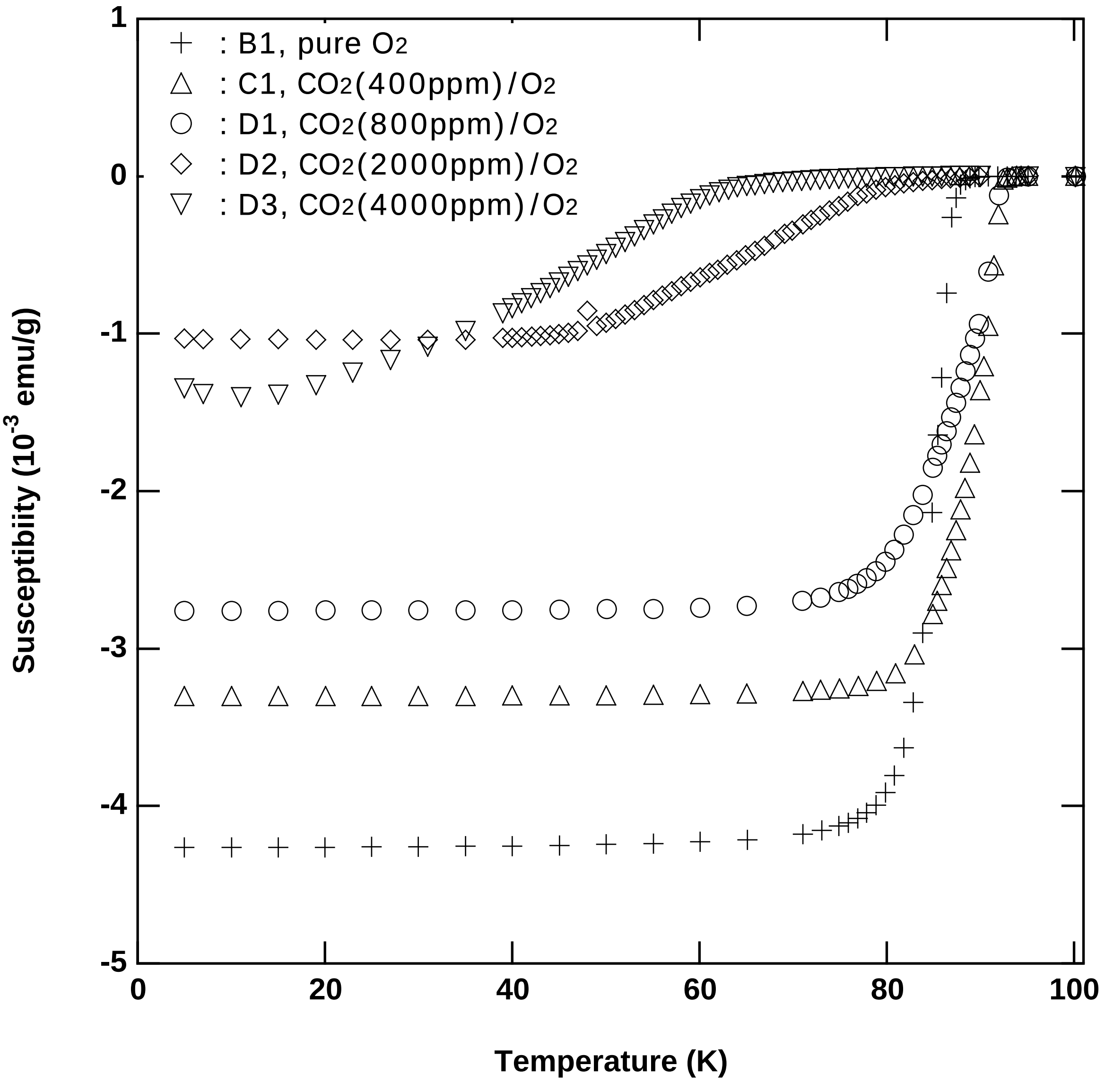

Fig. 8 A.Yamamoto et al.

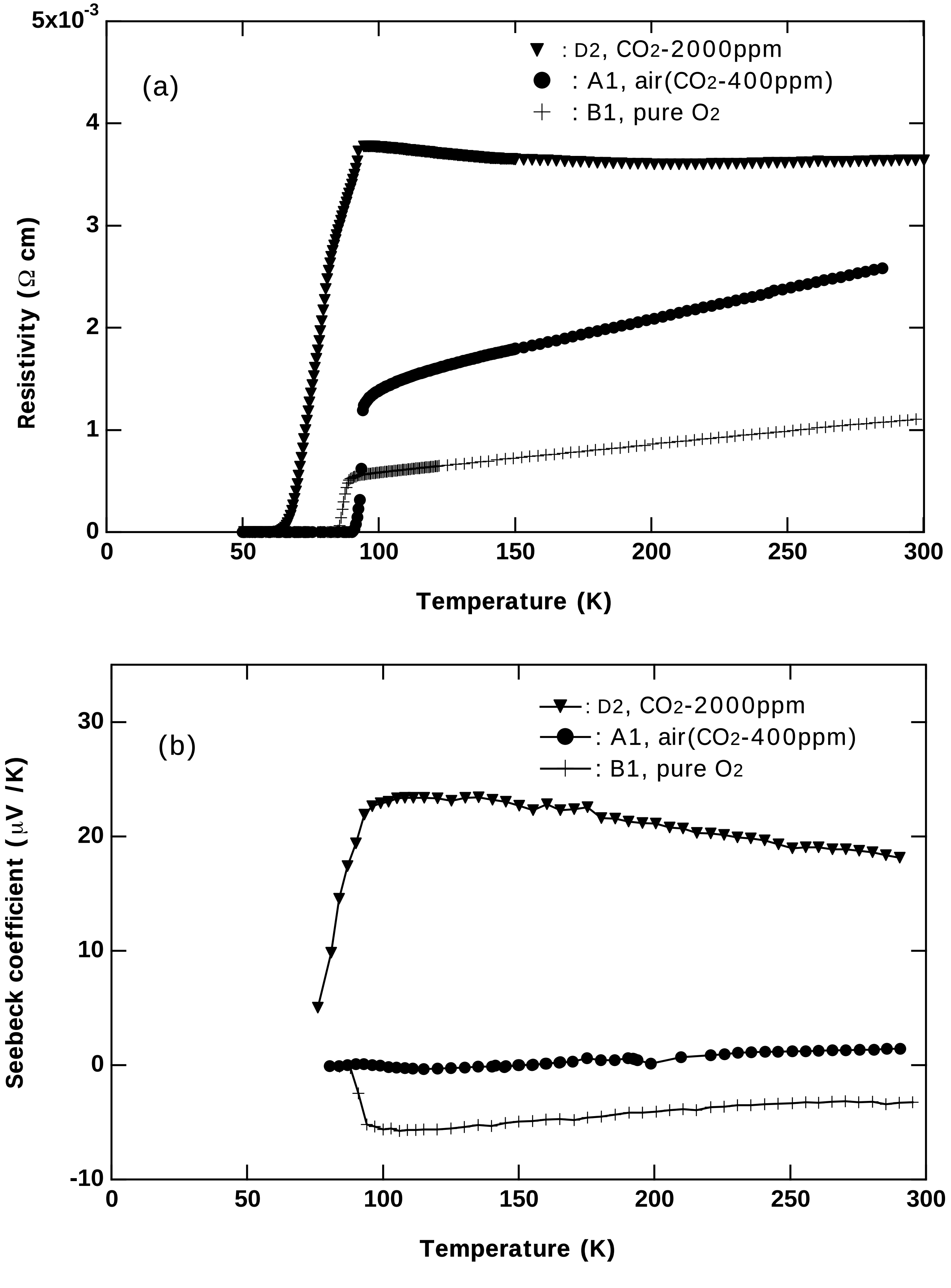

Fig. 9 A. Yamamoto et al.